\documentclass[amsfonts, amssymb, amsmath, aps, pra, superscriptaddress, english, floatfix, 10pt]{revtex4-2}
\usepackage[utf8]{inputenc}
\usepackage[colorlinks=true,citecolor=blue,linkcolor=blue,urlcolor=black]{hyperref}
\usepackage{physics}
\usepackage{dsfont}
\usepackage{orcidlink}

\newcommand{\average}[1]{\left\langle#1\right\rangle}

\begin{document}

\title{Entangled States are Harder to Transfer than Product States}

\author{Tony J. G. Apollaro \orcidlink{0000-0002-9324-9336}}
\email{tony.apollaro@um.edu.mt}
\affiliation{Department of Physics, University of Malta, Msida MSD 2080, Malta}

\author{Salvatore Lorenzo \orcidlink{0000-0002-0827-5549}}
\affiliation{Universit\`a degli Studi di Palermo, Dipartimento di Fisica e Chimica - Emilio Segr\`e, via Archirafi 36, I-90123 Palermo, Italy}

\author{Francesco Plastina \orcidlink{0000-0001-9615-8598}}
\affiliation{Dipartimento di Fisica, Universit\`a della Calabria, 87036 Arcavacata di Rende (CS), Italy}
\affiliation{INFN, gruppo collegato di Cosenza, 87036 Arcavacata di Rende (CS), Italy}

\author{Mirko Consiglio \orcidlink{0000-0001-8730-0206}}
\affiliation{Department of Physics, University of Malta, Msida MSD 2080, Malta}

\author{Karol \.{Z}yczkowski \orcidlink{0000-0002-0653-3639}}
\affiliation{Institute of Theoretical Physics,
Jagiellonian University, ul. {\L}ojasiewicza 11, 30--348 Krak\'ow, Poland}
\affiliation{Center for Theoretical Physics, Polish Academy of Sciences, Al. Lotnik\'{o}w 32/46, 02-668 Warszawa, Poland}

\begin{abstract}
    The distribution of entangled states is a key task of utmost importance for many quantum information processing protocols. A commonly adopted setup for distributing quantum states envisages the creation of the state in one location, which is then sent to (possibly different) distant receivers through some quantum channels. While it is undoubted and, perhaps, intuitively expected that the distribution of entangled quantum states is less efficient than that of product states, a thorough quantification of this inefficiency (namely, of the difference between the quantum-state transfer fidelity for entangled and factorized states) has not been performed. To this end, in this work, we consider $n$-independent amplitude-damping channels, acting in parallel, i.e., each, locally, on one part of an $n$-qubit state. We derive exact analytical results for the fidelity decrease, with respect to the case of product states, in the presence of entanglement in the initial state, for up to four qubits. Interestingly, we find that genuine multipartite entanglement has a more detrimental effect on the fidelity than two-qubit entanglement. Our results hint at the fact that, for larger $n$-qubit states, the difference in the average fidelity between product and entangled states increases with  increasing single-qubit fidelity, thus making the latter a less trustworthy figure of merit.
\end{abstract}

\maketitle

\section{Introduction}

Distributing entangled states among several distant recipients is a task of paramount importance in a variety of quantum-information processing protocols, ranging from $n$-party quantum key distribution~\cite{Hillery1999} to distributed quantum computing~\cite{Cuomo2020}. In many of these protocols, an $n$-partite entangled state is created at location $S$ (the sender's location), and its parts are distributed among $m\leq n$ receivers, generally at different locations (which we dub $R$, the receivers' location).

The special case of distributing a bipartite entangled state was already considered in the seminal paper by Bose on quantum-state transfer (QST), where the transfer protocol is employed to send (the state of) one party of a two-qubit Bell state to the opposite edge of a spin chain~\cite{Bose2003}. After this first instance, a considerable amount of research, both theoretical and experimental, has been performed in order to  improve the transfer performance and optimize the Bell state distribution protocol~\cite{Aspelmeyer2003,DiFranco2008b, Banchi2011, Apollaro2013a, Almeida2017,Streltsov2017, Wengerowsky2020}. Moreover, with the increasing  exploration (and exploitation) of  the fascinating realm of quantum correlations by quantum technological applications, the distribution of $n$-partite entangled states, with $n>2$, has become a very active research topic~\cite{Vieira2020,Apollaro2019a, Apollaro2020, Wang2020b, Mannalath2022}. 

At variance with the entanglement of two-qubits, which is the only system whose entanglement properties have been fully characterized both for pure and mixed states, for $n>2$ there are only a handful of closed, analytical results for the quantification of entanglement~\cite{Eltschka2014}, making the task of evaluating the efficiency of an entanglement distribution protocol very difficult to assess. 

Here, we address the distribution of an $n$-partite entangled state utilizing the fidelity between the sender's and the receivers' state as a figure of merit for the quality of the protocol. Although the fidelity is not a \textit{bona fide} tool to characterize quantum resources~\cite{Bina2014,Mandarino2014}, it is nevertheless widely employed in constructing entanglement witnesses following the idea that states close to an entangled state must be entangled as well~\cite{Guhne2009}. Hence, building on recent results~\cite{Lorenzo2022, Apollaro2022} reporting the fidelity of an $n$-qubits QST ($n$-QST) protocol for arbitrary quantum channels, we investigate the effect of the presence of entanglement on the $n$-QST fidelity, which is evaluated  when each qubit is subject to an independent $U(1)$-symmetric quantum channel, e.g., an amplitude-damping channel. In particular, we find that the presence of entanglement in the sender state is detrimental to the efficiency of the $n$-QST protocol. It is not surprising that independent quantum channels acting on $n$ qubits tend to destroy their quantum correlations, thus lowering the transmission fidelity. We are able to provide a quantification of the fidelity reduction as a function of  different entanglement monotones 
. In particular, we show that genuine multipartite entanglement, as quantified, e.g., by the three-tangle, has a more pronounced effect on lowering the $n$-QST fidelity than bipartite entanglement between two qubits, as quantified by the concurrence~\cite{Wootters1998}.

The paper is organised as follows: in Section~\ref{S_sec1}, we introduce our model and provide a brief recap of the $n$-QST fidelity; in Section~\ref{S_sec2}, we apply the developed formalism to the case of $n=2,3,4$ qubits; finally, in Section~\ref{S.D}, we draw our conclusions.

\section{\textit{n}-QST Fidelity for Independent Amplitude-Damping Channels}\label{S_sec1}

\begin{figure}
\centering
    \includegraphics[width=0.8\textwidth]
    {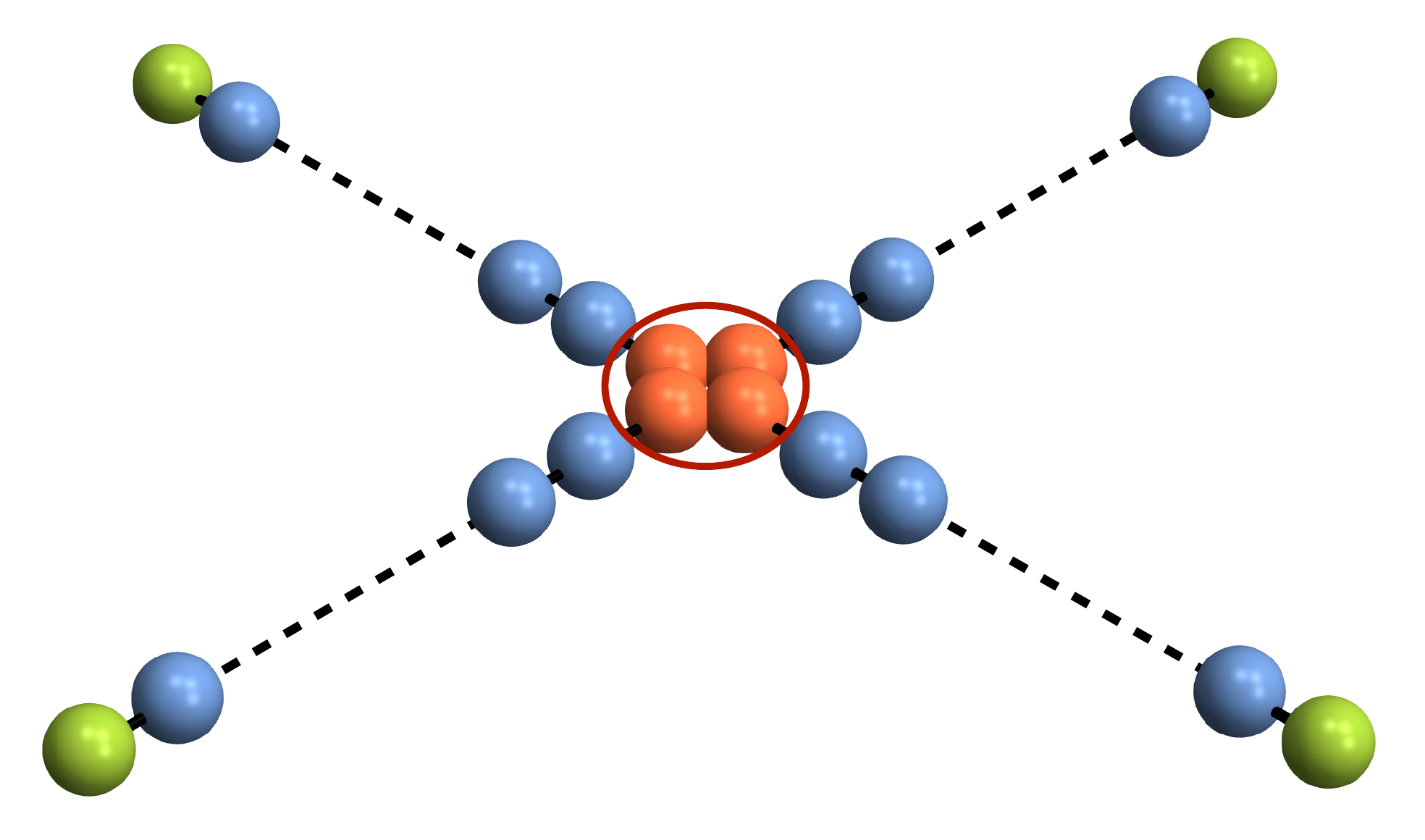}
    \caption{A quantum router. A dispatch center, encircled in red, creates an $n$-qubit entangled state (red spheres) with the aim to send each party to a different receiver (green spheres) along independent quantum channels (blue spheres).}
    \label{fig:model}
\end{figure}

Let us consider an $n$-qubit quantum-state transfer protocol as depicted in Figure~\ref{fig:model}. A sender, located at position $S$, prepares an $n$-qubit arbitrary state and wants to transfer each party to different receivers to which the sender is connected by different quantum channels. Without a loss of generality, let us assume the sender state to be a pure state, $\rho_S=\ketbra{\Psi}{\Psi}_n$. The state at the receivers' location reads
\begin{align}
    \label{eq:rec}
    \rho^R(t)=\left[\Phi_1\otimes\Phi_2\otimes \dots\otimes\Phi_n\right](t)\left(\rho_S\right).
\end{align}
The fidelity between the sender and the receivers' state is given by the Uhlmann–Jozsa fidelity~\cite{Jozsa1994}
\begin{align}
\label{E_UJfid}
F\left(\ket{\Psi},\rho(t)\right)=\bra{\Psi}\rho\ket{\Psi}~.
\end{align}
Expressing an arbitrary input state in the computational basis
\begin{align}
    \label{eq:input}
    \ket{\Psi}=\sum_{i=1}^{2^n}a_i\ket{i},
\end{align}
the elements of the  receivers' density matrix read (sum over repeated indexes is assumed)
\begin{eqnarray}
\label{eq:output}
\rho^R_{ij}=A_{ij}^{nm}\rho^S_{nm}\label{eq_rec}~
\end{eqnarray}
yielding the fidelity
\begin{align}
	\label{E.Fid_dis}
F\left(\ket{\Psi},\rho\right)=\sum_{ijnm=0}^{d-1}a_i^*a_ja_na_m^*A_{ij}^{nm},
\end{align}
where all of the amplitudes $a$ refer to the initial state of the sender.

For the case represented in Figure~\ref{fig:model}, the total map is given by the tensor products of $n$ independent maps as in Equation~\eqref{eq:rec}.
Hence, Equation~\eqref{eq:output} can be cast in the following form:~\cite{Bellomo2007}
\begin{align}
    \label{eq:indep_map_1}
 \rho^R_{i_1 i_2\dots i_n;j_1 j_2\dots j_n}=A_{i_1 i_2\dots i_n;j_1 j_2\dots j_n}^{p_1 p_2\dots p_n;q_1 q_2 \dots q_n}\rho^S_{p_1 p_2\dots p_n;q_1 q_2\dots q_n},
\end{align}
where $i,j,p,q=0,1$ and the corresponding subscript refers to the $i$'s qubit, with
\begin{align}
    \label{eq:indep_map}
    A_{i_1 i_2\dots i_n;j_1 j_2\dots j_n}^{p_1 p_2\dots p_n;q_1 q_2 \dots q_n}=A_{i_1;j_1}^{p_1;q_1}A_{i_2;j_2}^{p_2;q_2}\dots A_{i_n;j_n}^{p_n;q_n}.
\end{align}

Each $A$ in Equation~\eqref{eq:indep_map} comes from a single qubit map connecting the sender qubit $s_i$ and the receiver qubit $r_i$, which, for an $U(1)$-symmetric channel, can be expressed as
\begin{align}\label{E_1ex_map}
\begin{pmatrix}
\rho_{00}\\
\rho_{01}\\
\rho_{10}\\
\rho_{11}\\
\end{pmatrix}_{r_i}=
\begin{pmatrix}
1 &0&0&1-\left|f_{s_i}^{r_i}\right|^2\\
0&f_{s_i}^{r_i}&0&0\\
0&0&\left(f_{s_i}^{r_i}\right)^*&0\\
0&0&0&\left|f_{s_i}^{r_i}\right|^2
\end{pmatrix}
\begin{pmatrix}
\rho_{00}\\
\rho_{01}\\
\rho_{10}\\
\rho_{11}\\
\end{pmatrix}_{s_i},
\end{align}
where $f_{s_i}^{r_i}$ is the transition amplitude for the excitation initially on $s_i$ to reach $r_i$. 
A widely used $U(1)$-symmetric quantum channel is given by the so called $XXZ$ spin-$\frac{1}{2}$ Hamiltonian,
\begin{align}
    H=\sum_{i,j} J_{ij}\left(\sigma^x_i \sigma^x_j+\sigma^y_i \sigma^y_j\right)+\Delta_{ij} \sigma^z_i \sigma^z_j+h_i \sigma^z_i~
\end{align}
where $\sigma^{\alpha}_i $ ($\alpha=x,y,x$) are Pauli matrices and $i,j$ are the position indexes on an arbitrary $d$-dimensional lattice. Assuming that each quantum channel is fully polarized, for the sender state 
, the map $\Phi_i$ reduces to an amplitude-damping channel~\cite{Bose2007}. In particular, for $f_{s_i}^{r_i}=\mathds{1}$, the map $\Phi_i$ entails a SWAP operation. Therefore, our formalism  also describes entanglement swapping protocols via imperfect operations~\cite{Pan1998}. Finally, to express Equation~\eqref{eq:indep_map_1} in the form of Equation~\eqref{eq:output}, it is sufficient to express the bit strings in decimal notation.

By making use of Equations~\eqref{eq:output}, \eqref{eq:indep_map}, and \eqref{E_1ex_map}, it is straightforward to evaluate the average fidelity~\cite{Apollaro2020} of an arbitrary quantum state $\ket{\Psi}$
\begin{align}
\label{E_AvFid}
\average{F}=\frac{1}{\Omega}\int_{\Omega}d\Omega~F\left(\ket{\Psi},\rho(t)\right),
\end{align}
with $\Omega$ denoting the space of pure states and, with an abuse of notation, its volume and the measure of it 
.

An average with respect to an $n$ qubit system will be denoted as $\average{F_n}$.
In the case of $n$ independent channels, making use of the transition amplitude $f$ introduced in Equation~\eqref{E_1ex_map}, one arrives at the expression,
\begin{align}
	\label{E_F_ind}
	\average{F_n}=\frac{1}{2^n+1}+\frac{1}{2^n\left(2^n+1\right)}\left|1+ f\right|^{2n}~.
\end{align}
Notice that the average fidelity $\average{F_n} \leq \prod_{i=1}^{n}\average{F_1}$, with equality holding only for $f=0,1$.  While the left-hand side of the latter inequality gives the average over all possible pure input states,  its right-hand side, on the other hand, gives the average restricted to fully factorized states only, i.e., to product states of the form $\ket{\Psi}_n=\bigotimes_{i=1}^{n}\ket{\psi}_i$, thus not including the entangled states. Hence, we conclude that, when $n\geq2$, in the set of all pure input states, entangled states have a lower $n$-QST fidelity than the product state. In the next sections, we will provide a quantitative analysis for this intuitive observation.

\section{\textit{n}-QST Fidelity as a Function of Entanglement}\label{S_sec2}
This Section contains our main result, namely, that the presence of entanglement reduces the transfer fidelity. Below, we illustrate this idea separately for two, three, and four qubit transmissions. In particular, we will show that, in the presence of entanglement in the states to be sent, a {\it reduction factor} exists, which we dub $R_n$, such that the average fidelity for the QST of $n$-qubits can be generically written
$$ \langle F_n \rangle = \langle F_1 \rangle^n - {\cal E}_{n} \,  R_n.  $$
Here, $\langle F_1 \rangle^n$ gives the average fidelity for the transfer of factorized states (indeed, intuitively, qubits can be transferred one by one, in this case), so that the difference $ \langle F_n \rangle - \langle F_1 \rangle^n$ is entirely due to the fact that entangled states are possibly transferred. The coefficient ${\cal E}_{n}$ is an entanglement quantifier that changes with $n$. It is given by twice the square of concurrence for $n=2$, while for $n=3$ it is proportional to a linear combination of the invariant polynomials identifying the different classes of entangled states. Finally, the factor $R_n$ (defined below for the various cases) gives the weight of entanglement-induced fidelity decrease, and it also enters the fidelity averaged over specific entanglement classes of states (for $n=3,4$).

\subsection{Two Qubits}
Adopting the (Schmidt-)parametrization of two-qubits pure states in terms of their entanglement~\cite{Apollaro2015}, we write
\begin{align}\label{eq:2_qubit_gen}
    \ket{\Psi(s)}&=\sqrt{\frac{1 + s}{2}}\ket{00}+\sqrt{\frac{1 - s}{2}}\ket{11} 
\end{align}
where the parameter $s\in \left[-1,1\right]$ is related to the concurrence~\cite{Wootters1998} via $C=\sqrt{1-s^2}$, and every two-qubit pure state can be obtained from Equation~\eqref{eq:2_qubit_gen} via local, unitary operations $\ket{\Phi(s)}=U_1U_2\ket{\Psi(s)}$, with $U_i\in SU(2)$ acting on qubit $i=1,2$. Below, we obtain the average fidelity for a two-qubit QST protocol with independent channels as a function of the amount of entanglement of the sender state, which is invariant under local unitaries, and which we denote as $\average{\cdot}_{U^{\otimes 2}}$

This average fidelity  (which, to say it shortly, is averaged at fixed values of entanglement) reads
\begin{align}
    \label{eq:av_2F}
    \average{F_2}_{U^{\otimes 2}}
    =\frac{1}{36}\left(3+\left|f\right|^2+2\left|f\right|\cos\phi\right)^2-\frac{1}{18}\left(\left|f\right|^2+2\left|f\right|\cos\phi\right)\left(3-\left|f\right|^2+2\left|f\right|\cos\phi\right)C^2
\end{align}
where we expressed the complex transition amplitude $f$ as $f=|f| \, e^{i\phi}$. 

Following the procedure outlined by Bose~\cite{Bose2003}, in order to maximise Equation~\eqref{eq:av_2F}, one sets $\cos\phi=1$, which, physically, can be obtained, e.g., by a uniform magnetic field applied over the spin chain. Hence, the average two-qubit fidelity $F_2$ can be cast in the form
\begin{align}
    \label{eq:av_2F_1}
    \average{F_2}_{U^{\otimes 2}}=\frac{1}{36}\left(3+\left|f\right|^2+2\left|f\right|\right)^2-\frac{1}{18}\left(\left|f\right|^2+2\left|f\right|\right)\left(3-\left(\left|f\right|^2+2\left|f\right|\right)\right)C^2.
\end{align}
From Equation~\eqref{eq:av_2F_1}, since $0\leq \left|f\right| \leq 1$, one can readily appreciate that the more concurrence the sender's pure state contains, the lower the fidelity with the received state. Equation~\eqref{eq:av_2F_1} can also be rewritten as a function of the 
single-qubit QST average fidelity
\begin{align}
    \label{eq:1_avF}
    \average{F_1}=\frac{1}{6}\left(3+2 \left|f\right|+\left|f\right|^2\right),
\end{align}
to read
\begin{align}
    \label{eq:av_2F_2}
    \average{F_2}_{U^{\otimes 2}}=\average{F_1}^2-2\left(\average{F_1}-\frac{1}{2}\right)\left(1-\average{F_1}\right)C^2=\average{F_1}^2-2 R_2 C^2.
\end{align}
Again, as $\frac{1}{2}\leq \average{F_1} \leq 1$, the average fidelity $\average{F_2}$ decreases with the amount of concurrence of the sender state.

\begin{figure}[ht]
    \includegraphics[width=0.49\textwidth]{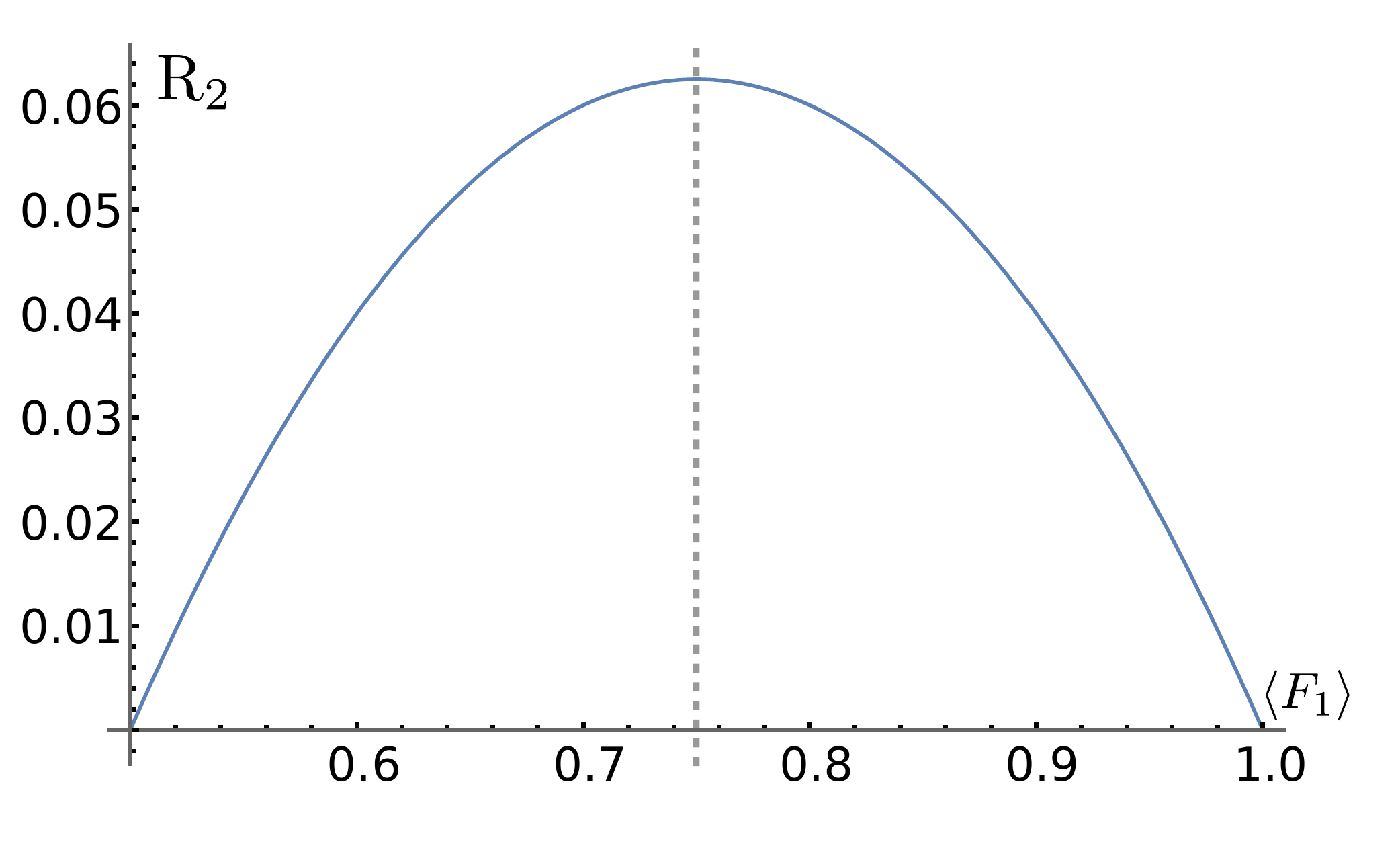}
       \includegraphics[width=0.49\textwidth]{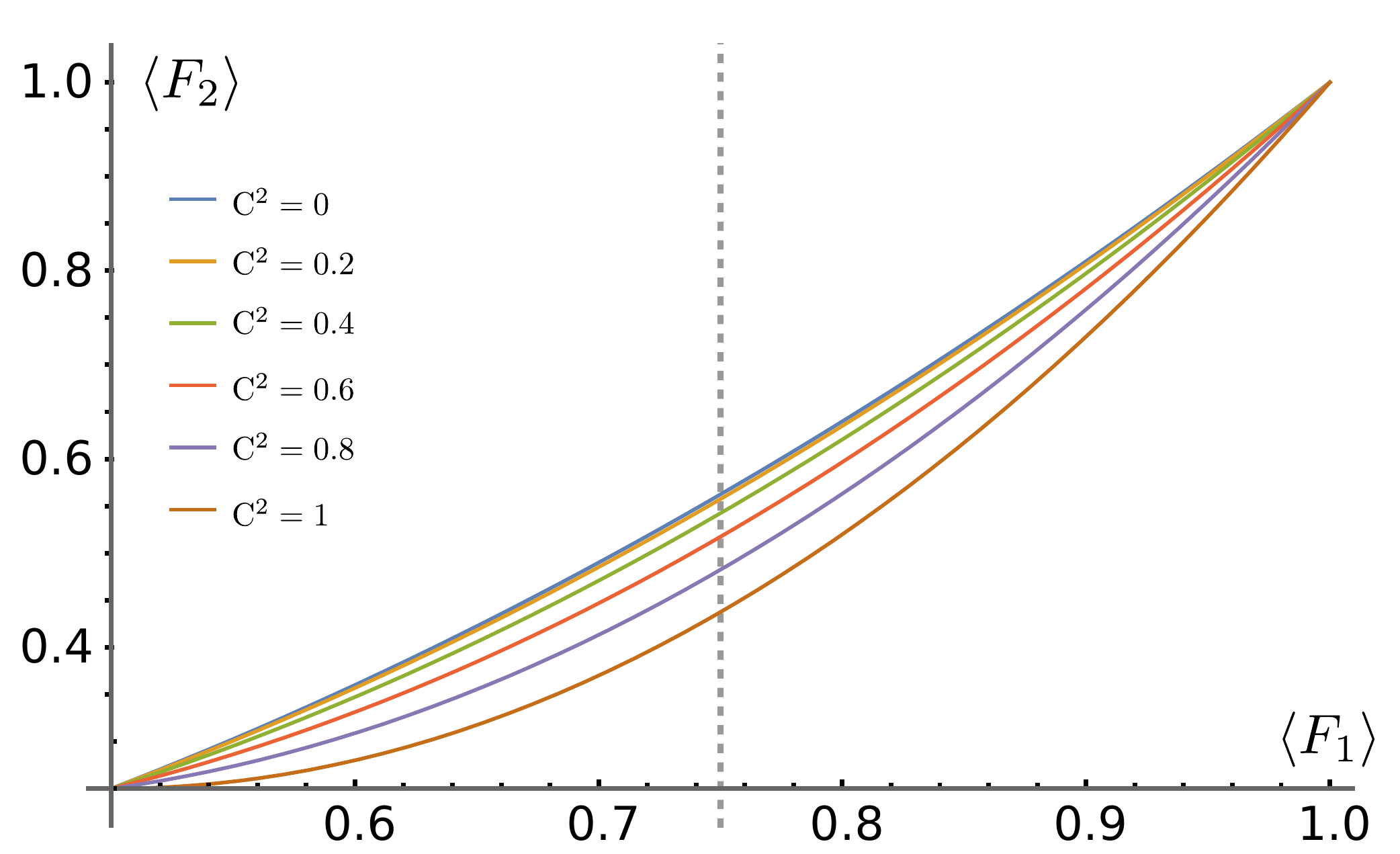}
    \caption{(left) Reduction factor \eqref{eq:redF_2} for entangled states of the 2-QST average fidelity $\average{F_2}$ \eqref{eq:av_2F_2} as a function of the 1-QST average fidelity $\average{F_1}$. The dotted, vertical line reports the maximum of $R_2$ attained at $\average{F}_1= 0.75$. (left) }
    \label{fig:rfactor_2}
\end{figure}

From Equation~\eqref{eq:av_2F_2}, we see that the average 2-QST fidelity is reduced in the presence of the squared concurrence by a factor of
\begin{align}
    \label{eq:redF_2}
    R_2=\left(\average{F_1}-\frac{1}{2}\right)\left(1-\average{F_1}\right)~,
\end{align}
which is reported in Figure~\ref{fig:rfactor_2} (left panel), together with the two-qubit average fidelity $\average{F_2}$, displayed for different values of the squared concurrence as a function of the one-qubit fidelity $\average{F_1}$ (right panel).

\subsection{Three Qubits}
Having obtained a quantitative expression giving the reduction in the transmission fidelity due to the presence of entanglement for two qubits, we move to the more intricate three qubit case in order to try and obtain similar relations.
\subsubsection{Three-qubit pure-state entanglement}

Let us now consider a system of three qubits $A, B$, and $C$.
A three-qubit pure state can be written in canonical form as~\cite{Acin2000}
\begin{equation}
\ket{\Psi}_{ABC}=\lambda_0 \ket{000}+\lambda_1 e^{i\phi}\ket{100}+\lambda_2 \ket{101}+\lambda_3 \ket{110}+\lambda_4 \ket{111},
\label{eq:3_canonical_form}
\end{equation}
where $\lambda_i\geq 0$, $0\leq \phi \leq \pi$, and the normalisation condition reads $\sum_{i} \lambda_i^2=1$. 

In terms of the coefficients of the state in Equation~\eqref{eq:3_canonical_form}, one can introduce five invariant polynomials, allowing to identify different entanglement classes~\cite{Enriquez2018a}:
\begin{subequations}
\begin{align}
    J_1&=\abs{\lambda_1\lambda_4 e^{i\phi}-\lambda_2\lambda_3}^2~,~    J_2=\lambda_0^2\lambda_2^2~,~
    J_3=\lambda_0^2\lambda_3^2 \\
    J_4&=\lambda_0^2\lambda_4^2~,~    J_5=\lambda_0^2\left(J_1+\lambda_2^2\lambda_3^2-\lambda_1^2\lambda_4^2\right)~.
\end{align}
\end{subequations}
The relation between invariant polynomials and entanglement measures is given by
\begin{align}
\label{eq_Ji_C}
    C_{jk}^2 = 4J_i,
\end{align}
for $i\neq j\neq k = 1,2, 3$, and where now, $\left(1,2,3 \right)=\left(A,B,C \right)$ holds on the LHS of Equation~\eqref{eq_Ji_C}~.
At variance with the two-qubit case, no single entanglement measure can capture genuine three-partite entanglement as three qubits can be entangled in two inequivalent ways~\cite{Du2000}.

One type of entanglement is quantified by the three-tangle~\cite{Coffman2000},
\begin{align}
    \tau_3^2 = 4J_4~,
\end{align}
while an inequivalent type of genuine multipartite entanglement (GME) is quantified by the so called GME concurrence, $C_{\text{GME}}$~\cite{Ma2011a}, defined, in terms of invariant polynomials, for a three-qubit pure state as:
\begin{align}
    \label{eq:3GME}
   C_{\text{GME}}
   = 4\left(\min\left\{J_2 + J_3, J_1 + J_3, J_1 + J_2\right\} + J_4 \right).
\end{align}


Hence, three qubit states can be sorted in the following entanglement classes~\cite{Du2000}:
\begin{itemize}
    \item \underline{Product states}. All $J_i=0$, resulting into $A-B-C$ (class 1). All entanglement measures vanish.
    \item \underline{Biseparable states}. All $J_i=0$ except i) $J_1$ for $A-BC$, ii) $J_2$ for $B-AC$, and iii) $J_3$ for $C-AB$ (class 2a). Only the concurrence for one single pair of qubits is different from zero.
    \item \underline{$W$-states}. $C_{\text{GME}}>0$ and  $\tau_3=0$
        \begin{enumerate}
            \item $J_4=0$ and $J_1 J_2+J_1J_3+J_2J_3=\sqrt{J_1J_2J_3}=\frac{J_5}{2}$ (class 3a).
            \item $J_4=0$ and $\sqrt{J_1J_2J_3}=\frac{J_5}{2}$ (class 4a).
        \end{enumerate}
    \item \underline{GHZ-states}. $C_{\text{GME}}>0$ and  $\tau_3>0$, with $5$ possible cases:
        \begin{enumerate}
            \item All $J_i=0$ except $J_4$ (class 2b).
            \item $J_1=J_2=J_5=0$, or $J_1=J_3=J_5=0$ or $J_2=J_3=J_5=0$ (class 3b).
            \item $J_2=J_5=0$ or $J_3=J_5=0$ (class 4b).
            \item $J_1 J_4+J_1 J_2+J_1 J_3=\sqrt{J_1J_2J_3}=\frac{J_5}{2}$ (class 4c).
            \item $\sqrt{J_1J_2J_3}=\frac{\left|J_5\right|}{2}$ and $\left(J_4+J_5\right)^2-4\left(J_1+J_4\right)\left(J_2+J_4\right)\left(J_3+J_4\right)=0$ (class 4d)~,
        \end{enumerate} 
\end{itemize}
where, with the notation $X-Y$, we indicate that subsystems $X$ and $Y$ do not share any type of entanglement.


Notably, for 3-qubit pure states, a monogamy relation exists between the amount of entanglement that can be shared among the parties~\cite{Coffman2000}
\begin{align}
    \label{eq:monogamy}
    C^2_{A|BC}=C^2_{AB}+C^2_{AC}+\tau_3^2~.
\end{align}
\subsubsection{Fidelity of 3-QST}

Here, we derive the fidelity of the QST of a tree-qubit pure state. First, we average the state in Equation~\eqref{eq:3_canonical_form} over single-qubit, local operations $U$, and use the notation $\average{\cdot}_{U^{\otimes 3}}$, in order to indicate the average fidelity at given values of (and, thus, as a function of) the entanglement quantifiers. 
Subsequently, utilizing the averages of the invariant polynomials obtained in Ref.~\cite{Enriquez2018a}, we derive the average fidelity of each three-qubit class and use the notation $\average{\cdot}$.

The 3-QST fidelity, expressed in terms of the single-qubit average, reads
\begin{align}
    \label{eq:fid_3}
  \average{F(\ket{\Psi},\rho_0)}_{U^{\otimes 3}}= \average{F_1}^3-8 \average{F_1}\left( \average{F_1}-\frac{1}{2}\right)\left(1-\average{F_1}\right) \left(J_1+J_2+J_3+\frac{3}{2}J_4\right).
\end{align}

Since  $\frac{1}{2}\leq \average{F_1}\leq 1$, and $0\leq J_i\leq \frac{1}{4}$, the second term on the right hand side of the latter equation is always negative, so that one sees at once that entanglement reduces the 3-QST fidelity. 

Introducing the reduction factor $R_3$, 
\begin{align}
    \label{eq:redF_3}
    R_3=\average{F_1}\left(\average{F_1}-\frac{1}{2}\right)\left(1-\average{F_1}\right)~,
\end{align}
we can write
\begin{align}
  \average{F(\ket{\Psi},\rho_0)}_{U^{\otimes 3}}= \average{F_1}^3-8 \, R_3 \, \left(J_1+J_2+J_3+\frac{3}{2}J_4\right).
\end{align}

Now, using the averages of the invariant polynomials obtained in Ref.~\cite{Enriquez2018a}, $\average{J_4}=\frac{1}{12}$ and $\average{J_k}=\frac{1}{24}$ ($k=1,2,3$), the average fidelity $\average{F_3}$ (with average taken over the full three qubit Hilbert space) is given by

\begin{align}
    \label{eq:fid_4}
   \average{F_3}= \average{F_1}^3-2 R_3
\end{align}

Here, we see that, at fixed single-particle average fidelity, entanglement  is responsible for a decrease in the 3-QST average fidelity by twice the reduction factor $R_3$.


In particular, the fidelities for the canonical states belonging to the different entanglement classes read
\begin{itemize}
    \item class 1 (product state) 
    \begin{align}
        \label{eq:fid_1a}
        &\average{F_\text{c1}}_{U^{\otimes 3}}=\average{F_1}^3\\
        &\average{F_\text{c1}}=\average{F_1}^3\
    \end{align}
    \item class 2a (biseparable states) 
    \begin{align}
        \label{eq:fid_2a}
        \average{F_\text{c2a}}_{U^{\otimes 3}}
        &=\average{F_1}^3-2 \average{F_1}\left( \average{F_1}-\frac{1}{2}\right)\left(1-\average{F_1}\right) C_{jk}^2\\
        \average{F_\text{c2a}}
        &=\average{F_1}^3-\frac{R_{3}}{3}
    \end{align}
    where $i\neq j\neq k=1,2,3$;
    \item class 2b (GHZ-states)
        \begin{align}
        \label{eq:fid_2b}
      \average{F_\text{c2b}}_{U^{\otimes 3}}
        &=\average{F_1}^3-3\average{F_1}\left( \average{F_1}-\frac{1}{2}\right)\left(1-\average{F_1}\right)\tau_3^2\\
        \average{F_\text{c2b}}
        &=\average{F_1}^3-R_{3}
    \end{align}
    \item class 3a ($J_4=0$ and $J_1 J_2+J_1J_3+J_2J_3=\sqrt{J_1J_2J_3}=\frac{J_5}{2}$)
    \begin{align}
    \label{eq:fid_3a}
   \average{F_\text{c3a}}_{U^{\otimes 3}}
   &=\average{F_1}^3-2\average{F_1}\left( \average{F_1}-\frac{1}{2}\right)\left(1-\average{F_1}\right) \left(C^2_{BC}+C^2_{AC}+C^2_{AB}\right)\nonumber\\
    \average{F_\text{c3a}}&=\average{F_1}^3-R_{3}
    \end{align}
%
    \item Class 3b ($J_1=J_2=J_5=0$ or $J_1=J_3=J_5=0$ or $J_2=J_3=J_5=0$)
     \begin{align}
        \label{eq:fid_3b}
        \average{F_\text{c3b}}_{U^{\otimes 3}}
        &=\average{F_1}^3- \average{F_1}\left( \average{F_1}-\frac{1}{2}\right)\left(1-\average{F_1}\right) \left(2C_{BC}^2+3\tau_3^2\right)\nonumber\\
       \average{F_\text{c3b}} &=\average{F_1}^3-\frac{4}{3}R_{3}
    \end{align}
%
    \item Class 4a $J_4=0$ and $\sqrt{J_1J_2J_3}=\frac{J_5}{2}$
    \begin{align}
        \label{eq:fid_4a}
   \average{F_\text{c4a}}_{U^{\otimes 3}}
   &=\average{F_1}^3-2\average{F_1}\left( \average{F_1}-\frac{1}{2}\right)\left(1-\average{F_1}\right) \left(C^2_{BC}+C^2_{AC}+C^2_{AB}\right)\nonumber\\
    \average{F_\text{c4a}}&=\average{F_1}^3-R_{3}
    \end{align}
%
    \item Class 4b ($J_2=J_5=0$ or $J_3=J_5=0$)
        \begin{align}
        \label{eq:fid_4b}
        \average{F_\text{c4b}}_{U^{\otimes 3}}
        &=\average{F_1}^3- \average{F_1}\left( \average{F_1}-\frac{1}{2}\right)\left(1-\average{F_1}\right) \left(2\left(C_{BC}^2+C_{AC}^2\right)+3\tau_3^2\right)\nonumber \\
         \average{F_\text{c4b}}&=\average{F_1}^3-\frac{5}{3}R_{3}
    \end{align}
    
    \item Class 4c ($J_1J_4 + J_1J_2 + J_1J_3 + J_2J_3 = \sqrt{J_1J_2J_3} = \frac{J_5}{2}$)
        \begin{align}
        \label{eq:fid_4c}
        \average{F_\text{c4c}}_{U^{\otimes 3}}
        &=\average{F_1}^3- \average{F_1}\left( \average{F_1}-\frac{1}{2}\right)\left(1-\average{F_1}\right) \left(2\left(C_{BC}^2+C_{AC}^2+C_{AB}^2\right)+3\tau_3^2\right) \nonumber \\
        \average{F_\text{c4c}}&=\average{F_1}^3-2R_{3}
        \end{align}
        
    \item Class 4d ($\sqrt{J_1J_2J_3}=\frac{\left|J_5\right|}{2}$ and $\left(J_4+J_5\right)^2-4\left(J_1+J_4\right)\left(J_2+J_4\right)\left(J_3+J_4\right)=0$)
        \begin{align}
        \label{eq:fid_4d}
        \average{F_\text{c4c}}_{U^{\otimes 3}}
        &=\average{F_1}^3- \average{F_1}\left( \average{F_1}-\frac{1}{2}\right)\left(1-\average{F_1}\right) \left(2\left(C_{BC}^2+C_{AC}^2+C_{AB}^2\right)+3\tau_3^2\right) \nonumber \\
        \average{F_\text{c4d}}&=\average{F_1}^3-2R_{3}
    \end{align}
\end{itemize}

\begin{figure}
    \includegraphics[width=0.49\textwidth]{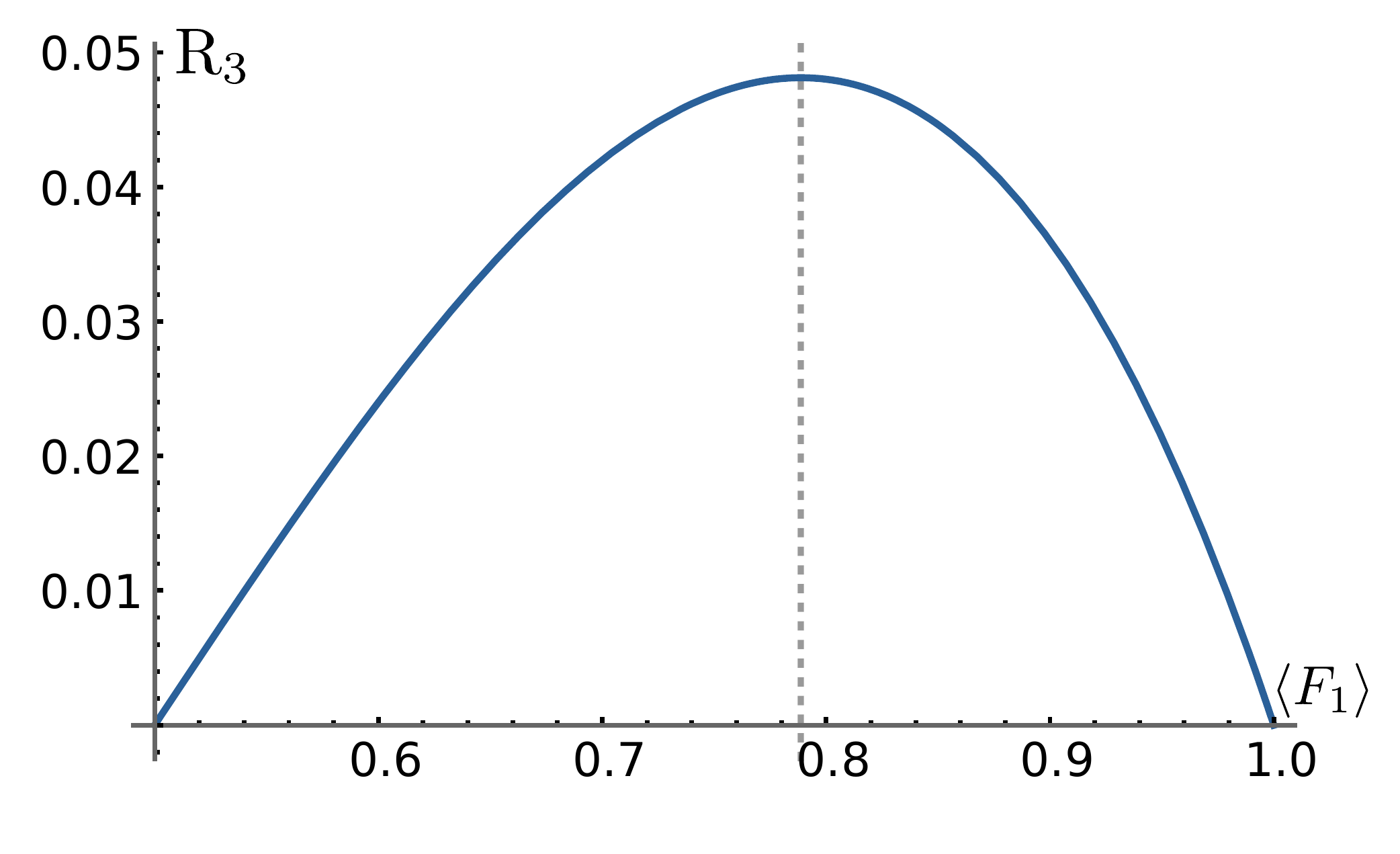}
        \includegraphics[width=0.49\textwidth]{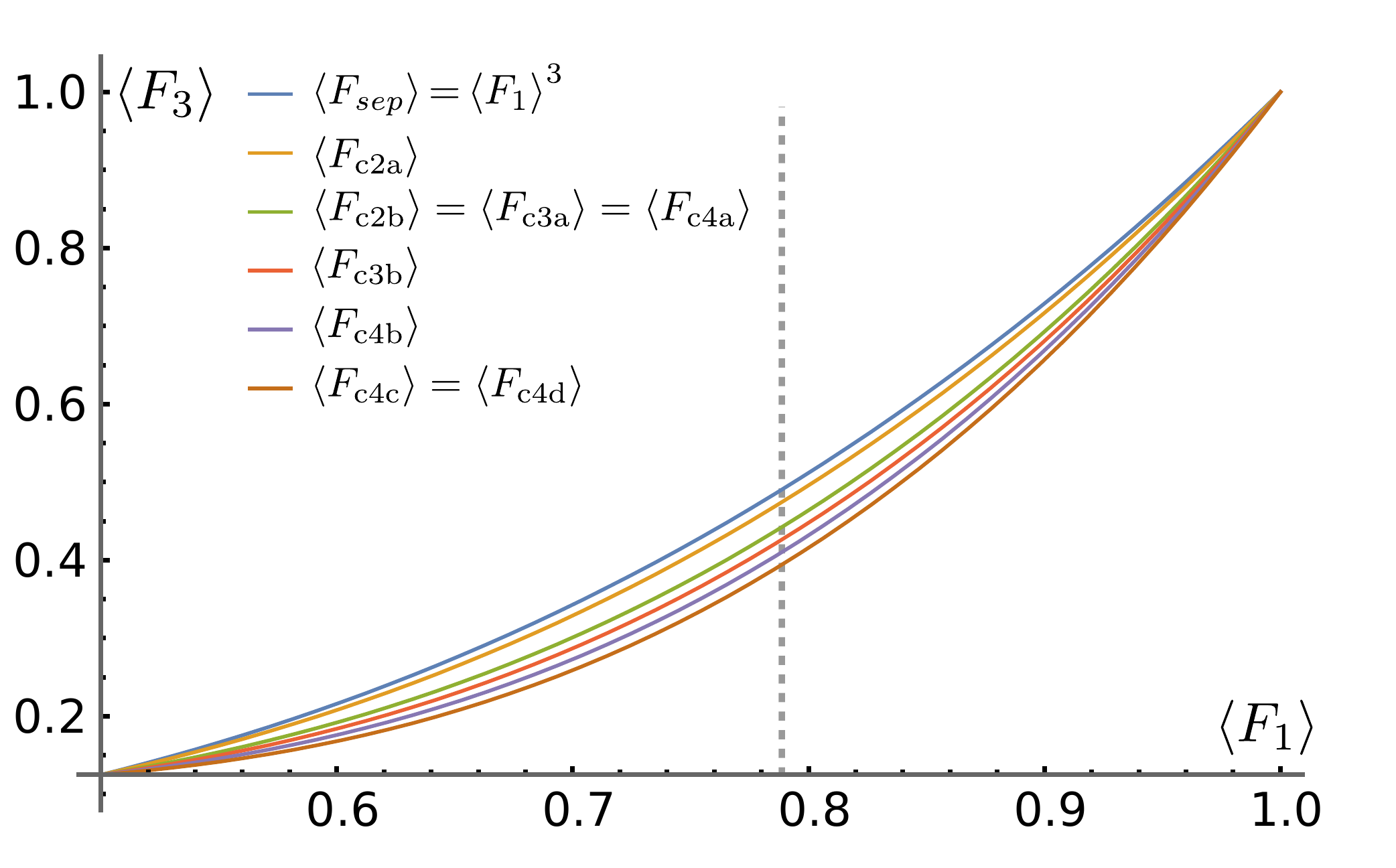}
    \caption{(left) Reduction factor for the average fidelity in the presence of entanglement as in Equation~\eqref{eq:redF_3}. (right) Average fidelity for the three-qubit classes. The dotted, vertical line is at $\average{F}_1=\frac{1}{2} \left(1+\frac{1}{\sqrt{3}}\right)\simeq 0.789$. 
    }
    \label{fig:rfactor_3}
\end{figure} 


Comparing Equation~\eqref{eq:fid_2a} with Equation~\eqref{eq:fid_2b}, it turns out that, for an equivalent amount of the entanglement monotone $C_{jk}^2$ and $\tau_3^2$, at fixed $\average{F_1}$ (or, equivalently, at a fixed transition amplitude $f$), the fidelity of the canonical state in class 2a is greater than that in class 2b. This is in line with our intuition that the more entangled a state is, the harder it is to achieve high fidelity in our parallel QST protocol, as shown in Figure~\ref{fig:rfactor_3} (right panel), where we plot the average fidelity for different three qubit classes. In the left panel of Figure~\ref{fig:rfactor_3} we report the reduction factor $R_3$ of Equation~\eqref{eq:redF_3} as a function of the single-particle average fidelity $\average{F_1}$.

From the above equations of the average fidelity of the three-qubit classes, we see that the average fidelity is decreased, with respect to the product state class, whenever there is two-qubit concurrence or genuine multipartite entanglement, both as $C_{\text{GME}}$ and as $\tau_3$. Moreover, per equal amount of squared two-qubit concurrence $C^2$ and genuine multipartite entanglement $C_\text{GME}$, the reducing factor is respectively 2 and 3 times $R_3$. As a consequence, we state that, at fixed amount of entanglement, GME states are harder to transfer than biseparable states.


\subsection{Four Qubits}

While for two and three qubits, the entanglement of pure states has been fully characterized, for four (or more) qubits there are infinitely many inequivalent entanglement classes~\cite{Du2000, Verstraete2002} under SLOCC operations (stochastic local operations and classical communication).

Here, we consider the fidelity of specific four-qubit states averaged over random local unitaries on each qubit. Whereas this does not account for all entangled states within a given class, as the group of stocastic local operations includes deterministic local operations, $SU(2)\subseteq SL(2, C)$ (with equality holding for pure states), the results nevertheless hint at the fact that the average fidelity decreases with the entanglement of the sender state and that the reduction factor depends on the type of entanglement contained in the state.

We will consider the three irreducibly balanced states~\cite{Osterloh2010}: the 4-qubits GHZ-state, the cluster, and $X_4$ 
:
\begin{subequations}
\label{eq_4qubits_irre}
\begin{align}
    \ket{\text{GHZ}_4}&=\frac{1}{\sqrt{2}}\left(\ket{0000}+\ket{1111}\right)\\
    \ket{\text{Cl}_4}&=\frac{1}{2}\left(\ket{0000}+\ket{0111}+\ket{1011}+\ket{1100} \right)\\
    \ket{X_4}&=\frac{1}{\sqrt{6}}\left(\sqrt{2}\ket{1111}+\ket{0001}+\ket{0010}+\ket{0100}+\ket{1000}\right)
\end{align}
\end{subequations}
and two additional entangled states: the product of two-Bell states and the 4-qubit $W$-state: 
\begin{subequations}
\label{eq_4qubits_add}
\begin{align}
    \ket{B_2}&=\ket{\Phi}_{12}\otimes \ket{\Phi}_{34}\\
    \ket{W_4}&=\frac{1}{2}\left(\ket{0001}+\ket{0010}+\ket{0100}+\ket{1000} \right)~.
\end{align}
\end{subequations}

\begin{figure}
\includegraphics[width=0.49\textwidth]{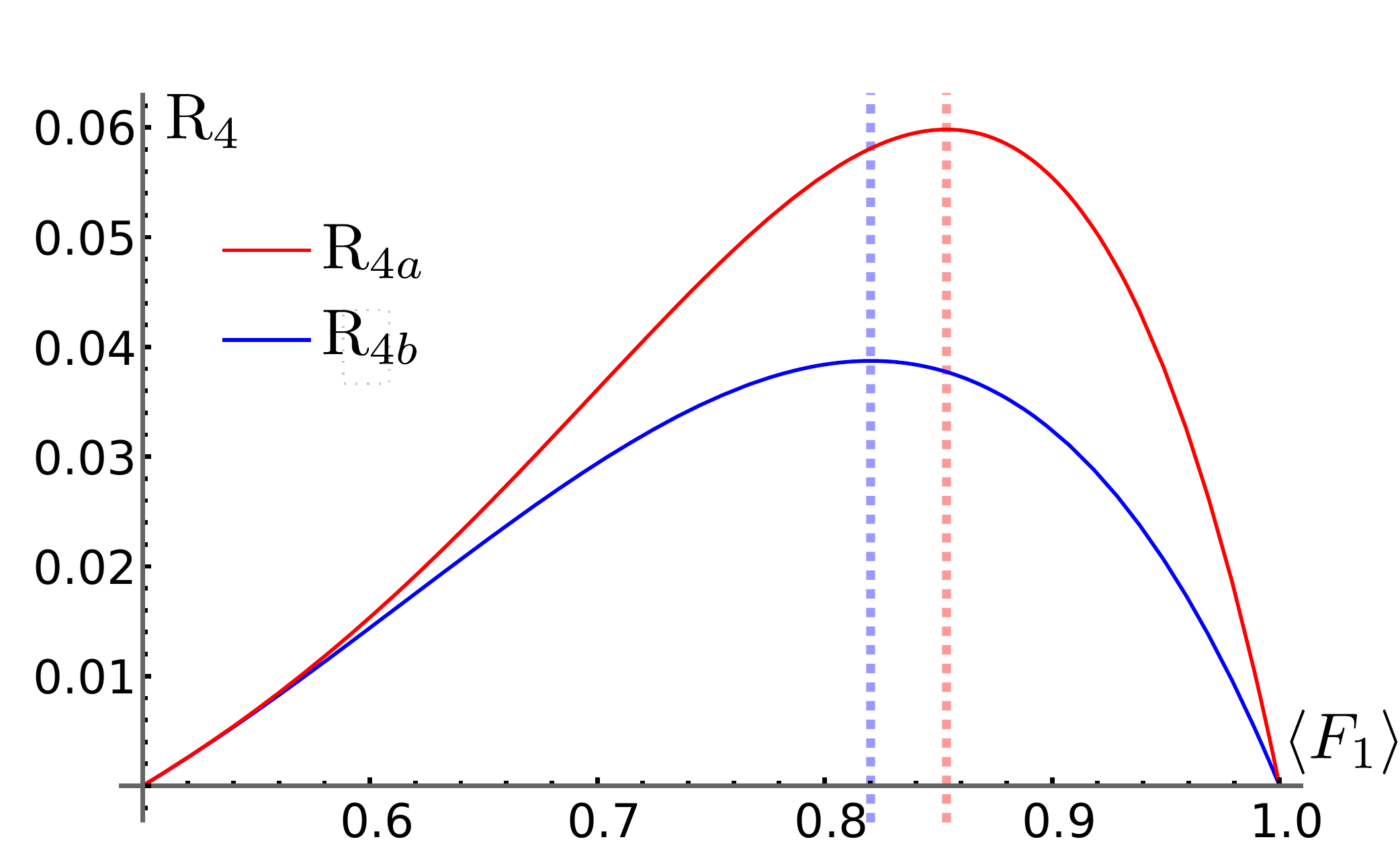}
    \includegraphics[width=0.49\textwidth]{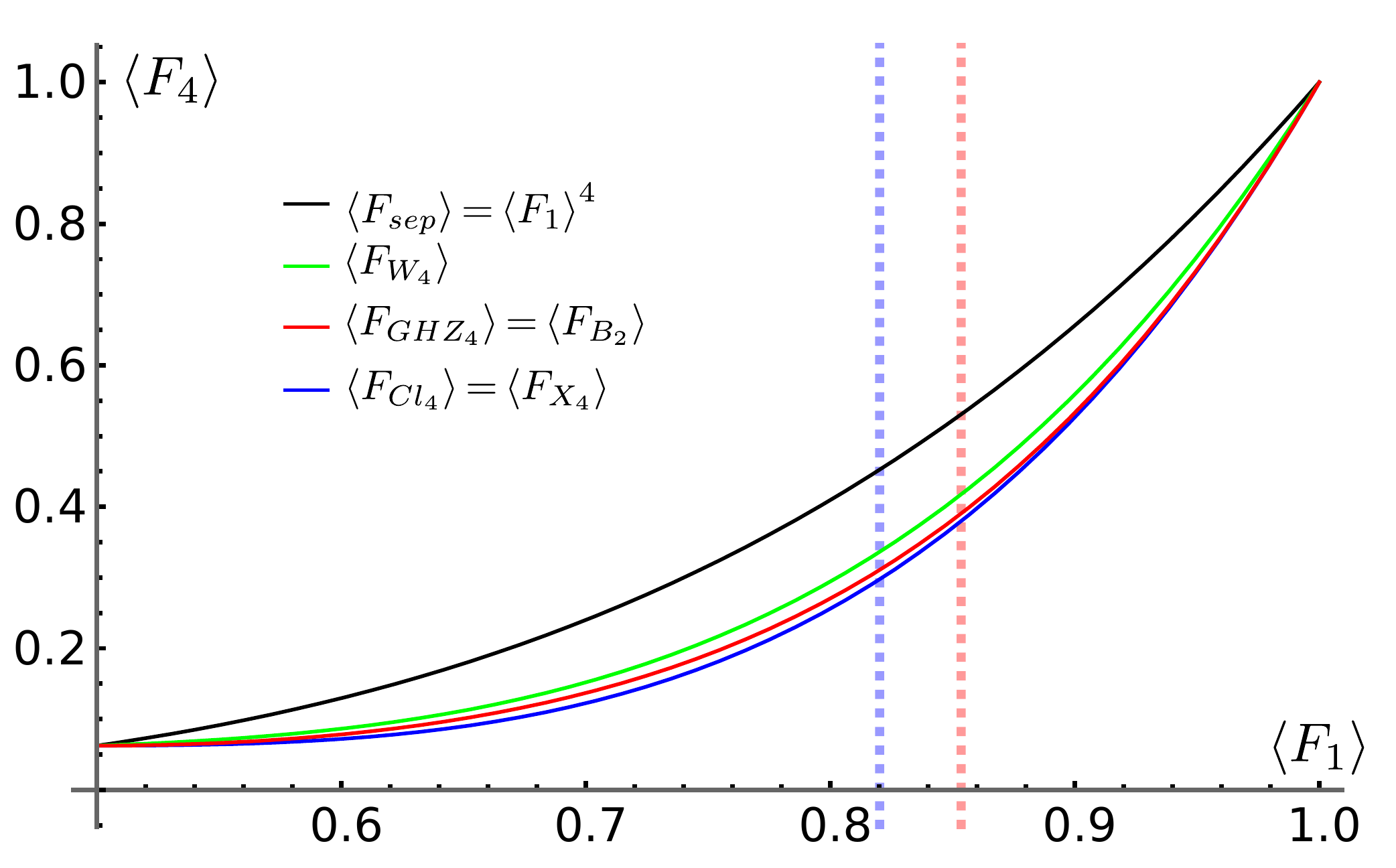}
    \caption{(left) Reduction factor for the average fidelity in the presence of entanglement as in Equations~\eqref{eq_red_4}. (right) Average fidelity for the entangled classes as reported in Equations~\eqref{eq_4qubits_irre_F}. The blue and red dotted, vertical lines are, respectively, at $\average{F}_1=0.82$ and $\average{F}_1=0.85$.}
    \label{fig:avg_fid_4}
\end{figure} 

The average fidelities of the states reported in Equations~\eqref{eq_4qubits_irre} and~\eqref{eq_4qubits_add}, expressed in terms of 1-QST average fidelity, read
\begin{subequations}
	    \label{eq_4qubits_irre_F}
\begin{align}
    &\average{F_{\text{GHZ}_4}}_{U^{\otimes 4}}=\average{F_{B_2}}_{U^{\otimes 4}}=\average{F_1}^4-2 \average{F_1}\left( \average{F_1}-\frac{1}{2}\right)\left(1-\average{F_1}\right)\left(1-3\average{F}+4\average{F}^2\right)\\
    &\average{F_{\text{Cl}_4}}_{U^{\otimes 4}}=\average{F_{X_4}}_{U^{\otimes 4}}=\average{F_1}^4-4\average{F_1}^2\left( \average{F_1}-\frac{1}{2}\right)\left(1-\average{F_1}\right)\\
    &\average{F_{W_4}}_{U^{\otimes 4}}=\average{F_1}^4-3\average{F_1}^2\left( \average{F_1}-\frac{1}{2}\right)\left(1-\average{F_1}\right)~.
\end{align}
\end{subequations}
Notice that the reduction factor for the states $\ket{Cl_4},\ket{X_4},\ket{W_4}$ is the same, although with different weights, and differs from the reduction factor for the states 
$\ket{GHZ_4}, \ket{B_2}$, reading, respectively,
\begin{subequations}
\label{eq_red_4}
\begin{align}
   R_{4a}&= \average{F_1}\left( \average{F_1}-\frac{1}{2}\right)\left(1-\average{F_1}\right)\left(1-3\average{F}+4\average{F}^2\right)\\
   R_{4b}&=\average{F_1}^2\left( \average{F_1}-\frac{1}{2}\right)\left(1-\average{F_1}\right)~.
\end{align}
\end{subequations}

A possible reason may be that, if one considers the four-tangle as an entanglement measure, although it is not a measure of genuine multipartite entanglement, the first set of state has zero four-tangle, whereas for the second one it is non-zero. In Figure~\ref{fig:avg_fid_4} (left panel) we report the reduction factors in Equation~\eqref{eq_red_4}, while in the right panel we report the average fidelity of Equation~\eqref{eq_4qubits_irre_F}.

\section{Discussion}\label{S.D}

We have shown that the QST of an entangled $n\geq 2$ quantum state across parallel, independent $U(1)$-symmetric quantum channels, as, e.g., embodied by an $XXZ$ spin-$\frac{1}{2}$ Hamiltonian, leads to a lower average fidelity than that of the QST of a product state at fixed one-qubit QST average fidelity, or, equivalently, at fixed transition amplitude. For the case of $n=2$, we have expressed the average fidelity reduction in terms of the squared concurrence times a reduction factor. Similarly, for $n=3$, we obtained that the presence of entanglement, both bipartite and multipartite, has a detrimental effect on the average fidelity. In particular, we obtained that the reduction factor has a greater weight in the presence of genuine three-partite entanglement, i.e., three-tangle and GME concurrence, than in the presence of two-qubit squared concurrence for specific canonical classes of the three-qubit pure state. Finally, we have considered specific cases of 4-qubit entangled states, which, again, result in an average fidelity reduction due to the presence of entanglement in the initial state.

Our work clearly shows that for entanglement distribution in a routing configuration, where parties are sent over independent quantum channels, the single-qubit average fidelity is not a reliable figure of merit. This calls for more investigations into the properties of quantum channels able to faithfully distribute multipartite entangled states.

\vspace{6pt}

\section*{Acknowledgements}

TJGA acknowledges funding through the IPAS+ (Internationalisation Partnership Awards Scheme +) QUEST project by the MCST (The Malta Council for Science \& Technology). MC acknowledges funding from the Tertiary Education Scholarships Scheme and from the Project QVAQT financed by the Malta Council for Science \& Technology, for and on behalf of the Foundation for Science and Technology, through the FUSION: R\&I Research Excellence Programme REP-2022-003.
K{\.Z} acknowledges support by Narodowe Centrum Nauki 
under the Quantera project number 2021/03/Y/ST2/00193 
and by the Foundation for Polish Science 
under the Team-Net project POIR.04.04.00-00-17C1/18-00.
SL acknowledges support by MUR under PRIN Project No. 2017 SRN-BRK QUSHIP


\begin{thebibliography}{0}%
\makeatletter
\providecommand \@ifxundefined [1]{%
 \@ifx{#1\undefined}
}%
\providecommand \@ifnum [1]{%
 \ifnum #1\expandafter \@firstoftwo
 \else \expandafter \@secondoftwo
 \fi
}%
\providecommand \@ifx [1]{%
 \ifx #1\expandafter \@firstoftwo
 \else \expandafter \@secondoftwo
 \fi
}%
\providecommand \natexlab [1]{#1}%
\providecommand \enquote  [1]{``#1''}%
\providecommand \bibnamefont  [1]{#1}%
\providecommand \bibfnamefont [1]{#1}%
\providecommand \citenamefont [1]{#1}%
\providecommand \href@noop [0]{\@secondoftwo}%
\providecommand \href [0]{\begingroup \@sanitize@url \@href}%
\providecommand \@href[1]{\@@startlink{#1}\@@href}%
\providecommand \@@href[1]{\endgroup#1\@@endlink}%
\providecommand \@sanitize@url [0]{\catcode `\\12\catcode `\$12\catcode
  `\&12\catcode `\#12\catcode `\^12\catcode `\_12\catcode `\%12\relax}%
\providecommand \@@startlink[1]{}%
\providecommand \@@endlink[0]{}%
\providecommand \url  [0]{\begingroup\@sanitize@url \@url }%
\providecommand \@url [1]{\endgroup\@href {#1}{\urlprefix }}%
\providecommand \urlprefix  [0]{URL }%
\providecommand \Eprint [0]{\href }%
\providecommand \doibase [0]{https://doi.org/}%
\providecommand \selectlanguage [0]{\@gobble}%
\providecommand \bibinfo  [0]{\@secondoftwo}%
\providecommand \bibfield  [0]{\@secondoftwo}%
\providecommand \translation [1]{[#1]}%
\providecommand \BibitemOpen [0]{}%
\providecommand \bibitemStop [0]{}%
\providecommand \bibitemNoStop [0]{.\EOS\space}%
\providecommand \EOS [0]{\spacefactor3000\relax}%
\providecommand \BibitemShut  [1]{\csname bibitem#1\endcsname}%
\let\auto@bib@innerbib\@empty
\end{thebibliography}%


\begin{thebibliography}{999}

\bibitem[Hillery \em{et~al.}(1999)Hillery, Bu{\v{z}}ek, and
  Berthiaume]{Hillery1999}
Hillery, M.; Bu{\v{z}}ek, V.; Berthiaume, A.
\newblock {Quantum secret sharing}.
\newblock {\em Phys. Rev. A} {\bf 1999}, {\em 59},~1829--1834.
\newblock {\url{https://doi.org/10.1103/PhysRevA.59.1829}}.

\bibitem[Cuomo \em{et~al.}(2020)Cuomo, Caleffi, and Cacciapuoti]{Cuomo2020}
Cuomo, D.; Caleffi, M.; Cacciapuoti, A.S.
\newblock {Towards a distributed quantum computing ecosystem}.
\newblock {\em IET Quantum Commun.} {\bf 2020}, {\em 1},~3--8,
  \href{http://xxx.lanl.gov/abs/2002.11808}{{\normalfont [2002.11808]}}.
\newblock {\url{https://doi.org/10.1049/iet-qtc.2020.0002}}.

\bibitem[Bose(2003)]{Bose2003}
Bose, S.
\newblock {Quantum Communication through an Unmodulated Spin Chain}.
\newblock {\em Phys. Rev. Lett.} {\bf 2003}, {\em 91},~1--4.
\newblock {\url{https://doi.org/10.1103/PhysRevLett.91.207901}}.

\bibitem[Aspelmeyer \em{et~al.}(2003)Aspelmeyer, B{\"{o}}hm, Gyatso, Jennewein,
  Kaltenbaek, Lindenthal, Molina-Terriza, Poppe, Resch, Taraba, Ursin, Walther,
  and Zeilinger]{Aspelmeyer2003}
Aspelmeyer, M.; B{\"{o}}hm, H.R.; Gyatso, T.; Jennewein, T.; Kaltenbaek, R.;
  Lindenthal, M.; Molina-Terriza, G.; Poppe, A.; Resch, K.; Taraba, M.;  et~al.
\newblock {Long-distance free-space distribution of quantum entanglement.}
\newblock {\em Science (New York, N.Y.)} {\bf 2003}, {\em 301},~621--3.
\newblock {\url{https://doi.org/10.1126/science.1085593}}.

\bibitem[{Di Franco} \em{et~al.}(2008){Di Franco}, Paternostro, and
  Kim]{DiFranco2008b}
{Di Franco}, C.; Paternostro, M.; Kim, M.
\newblock {Nested entangled states for distributed quantum channels}.
\newblock {\em Phys. Rev. A} {\bf 2008}, {\em 77},~1--4.
\newblock {\url{https://doi.org/10.1103/PhysRevA.77.020303}}.

\bibitem[Banchi \em{et~al.}(2011)Banchi, Bayat, Verrucchi, and
  Bose]{Banchi2011}
Banchi, L.; Bayat, A.; Verrucchi, P.; Bose, S.
\newblock {Nonperturbative Entangling Gates between Distant Qubits Using
  Uniform Cold Atom Chains}.
\newblock {\em Phys. Rev. Lett.} {\bf 2011}, {\em 106},~140501.
\newblock {\url{https://doi.org/10.1103/PhysRevLett.106.140501}}.

\bibitem[Apollaro \em{et~al.}(2013)Apollaro, Lorenzo, and
  Plastina]{Apollaro2013a}
Apollaro, T.J.G.; Lorenzo, S.; Plastina, F.
\newblock {Transport of quantum correlations across a spin chain}.
\newblock {\em Int. J. Mod. Phys. B} {\bf 2013}, {\em
  27},~1345035.
\newblock {\url{https://doi.org/10.1142/S0217979213450355}}.

\bibitem[Almeida \em{et~al.}(2017)Almeida, de~Moura, Apollaro, and
  Lyra]{Almeida2017}
Almeida, G.M.A.; de~Moura, F.A.B.F.; Apollaro, T.J.G.; Lyra, M.L.
\newblock Disorder-assisted distribution of entanglement in $XY$ spin chains.
\newblock {\em Phys. Rev. A} {\bf 2017}, {\em 96},~032315.
\newblock {\url{https://doi.org/10.1103/PhysRevA.96.032315}}.

\bibitem[Streltsov \em{et~al.}(2017)Streltsov, Kampermann, and
  Bru{\ss}]{Streltsov2017}
Streltsov, A.; Kampermann, H.; Bru{\ss}, D.
\newblock Entanglement Distribution and Quantum Discord. In {\em Lectures on
  General Quantum Correlations and their Applications}; Fanchini, F.F.;
  Soares~Pinto, D.d.O.; Adesso, G., Eds.; Springer International Publishing:
  Cham, Switzerland,  2017; pp. 217--230.
\newblock {\url{https://doi.org/10.1007/978-3-319-53412-1_10}}.

\bibitem[Wengerowsky \em{et~al.}(2020)Wengerowsky, Joshi, Steinlechner, Zichi,
  Liu, Scheidl, Dobrovolskiy, van~der Molen, Los, Zwiller, Versteegh, Mura,
  Calonico, Inguscio, Zeilinger, Xuereb, and Ursin]{Wengerowsky2020}
Wengerowsky, S.; Joshi, S.K.; Steinlechner, F.; Zichi, J.R.; Liu, B.; Scheidl,
  T.; Dobrovolskiy, S.M.; van~der Molen, R.; Los, J.W.; Zwiller, V.;  et~al.
\newblock {Passively stable distribution of polarisation entanglement over 192
  km of deployed optical fibre}.
\newblock {\em npj Quantum Inf.} {\bf 2020}, {\em 6},~1--5,
  \href{http://xxx.lanl.gov/abs/1907.04864}{{\normalfont [1907.04864]}}.
\newblock {\url{https://doi.org/10.1038/s41534-019-0238-8}}.

\bibitem[Vieira and Rigolin(2020)]{Vieira2020}
Vieira, R.; Rigolin, G.
\newblock {Almost perfect transmission of multipartite entanglement through
  disordered and noisy spin chains}.
\newblock {\em Phys. Lett. A} {\bf 2020}, {\em 384},~126536.
\newblock {\url{https://doi.org/10.1016/j.physleta.2020.126536}}.

\bibitem[Apollaro \em{et~al.}(2019)Apollaro, Almeida, Lorenzo, Ferraro, and
  Paganelli]{Apollaro2019a}
Apollaro, T.J.G.; Almeida, G.M.A.; Lorenzo, S.; Ferraro, A.; Paganelli, S.
\newblock {Spin chains for two-qubit teleportation}.
\newblock {\em Phys. Rev. A} {\bf 2019}, {\em 100},~052308.
\newblock {\url{https://doi.org/10.1103/PhysRevA.100.052308}}.

\bibitem[Apollaro \em{et~al.}(2020)Apollaro, Sanavio, Chetcuti, and
  Lorenzo]{Apollaro2020}
Apollaro, T.J.; Sanavio, C.; Chetcuti, W.J.; Lorenzo, S.
\newblock {Multipartite entanglement transfer in spin chains}.
\newblock {\em Phys. Lett. A} {\bf 2020}, {\em 384},~126306,
  \href{http://xxx.lanl.gov/abs/2001.03529}{{\normalfont [2001.03529]}}.
\newblock {\url{https://doi.org/10.1016/j.physleta.2020.126306}}.

\bibitem[Wang \em{et~al.}(2020)Wang, Xiang, Kang, Han, Liu, He, Gong, Su, and
  Peng]{Wang2020b}
Wang, M.; Xiang, Y.; Kang, H.; Han, D.; Liu, Y.; He, Q.; Gong, Q.; Su, X.;
  Peng, K.
\newblock {Deterministic Distribution of Multipartite Entanglement and Steering
  in a Quantum Network by Separable States}.
\newblock {\em Phys. Rev. Lett.} {\bf 2020}, {\em 125},~6388--6396,
  \href{http://xxx.lanl.gov/abs/2101.01422}{{\normalfont [2101.01422]}}.
\newblock {\url{https://doi.org/10.1103/PhysRevLett.125.260506}}.

\bibitem[Mannalath and Pathak(2022)]{Mannalath2022}
Mannalath, V.; Pathak, A.
\newblock Multiparty Entanglement Routing in Quantum Networks,  2022.
\newblock {\url{https://doi.org/10.48550/ARXIV.2211.06690}}.

\bibitem[Eltschka and Siewert(2014)]{Eltschka2014}
Eltschka, C.; Siewert, J.
\newblock {Quantifying entanglement resources}.
\newblock {\em J. Phys. Math. Theor.} {\bf 2014},
  {\em 47},~424005.
\newblock {\url{https://doi.org/10.1088/1751-8113/47/42/424005}}.

\bibitem[Bina \em{et~al.}(2014)Bina, Mandarino, Olivares, and Paris]{Bina2014}
Bina, M.; Mandarino, A.; Olivares, S.; Paris, M.G.A.
\newblock {Drawbacks of the use of fidelity to assess quantum resources}.
\newblock {\em Phys. Rev. A} {\bf 2014}, {\em 89},~012305,
  \href{http://xxx.lanl.gov/abs/1309.5325}{{\normalfont [1309.5325]}}.
\newblock {\url{https://doi.org/10.1103/PhysRevA.89.012305}}.

\bibitem[Mandarino \em{et~al.}(2014)Mandarino, Bina, Olivares, and
  Paris]{Mandarino2014}
Mandarino, A.; Bina, M.; Olivares, S.; Paris, M.G.
\newblock {About the use of fidelity in continuous variable systems}.
\newblock {\em Int. J. Quantum Inf.} {\bf 2014}, {\em
  12},~1--9,  \href{http://xxx.lanl.gov/abs/1402.0976}{{\normalfont
  [1402.0976]}}.
\newblock {\url{https://doi.org/10.1142/S0219749914610152}}.

\bibitem[G{\"{u}}hne and Toth(2008)]{Guhne2009}
G{\"{u}}hne, O.; Toth, G.
\newblock {Entanglement detection}.
\newblock {\em Phys. Rep.} {\bf 2008}, {\em 474},~1--75,
  \href{http://xxx.lanl.gov/abs/0811.2803}{{\normalfont [0811.2803]}}.
\newblock {\url{https://doi.org/10.1016/j.physrep.2009.02.004}}.

\bibitem[Lorenzo \em{et~al.}(2022)Lorenzo, Plastina, Consiglio, and
  Apollaro]{Lorenzo2022}
Lorenzo, S.; Plastina, F.; Consiglio, M.; Apollaro, T.J.G.
\newblock Quantum Map Approach to Entanglement Transfer and Generation in Spin
  Chains. In {\em Entanglement in Spin Chains: From Theory to Quantum
  Technology Applications}; Bayat, A.; Bose, S.; Johannesson, H., Eds.;
  Springer International Publishing: Cham, Switzerland,  2022; pp. 321--340.
\newblock {\url{https://doi.org/10.1007/978-3-031-03998-0_12}}.

\bibitem[Apollaro \em{et~al.}(2022)Apollaro, Lorenzo, Plastina, Consiglio, and
  {\.{Z}}yczkowski]{Apollaro2022}
Apollaro, T.J.G.; Lorenzo, S.; Plastina, F.; Consiglio, M.; {\.{Z}}yczkowski,
  K.
\newblock {Quantum transfer of interacting qubits}.
\newblock {\em New J. Phys.} {\bf 2022}, {\em 24},~083025,
  \href{http://xxx.lanl.gov/abs/2205.01579}{{\normalfont [2205.01579]}}.
\newblock {\url{https://doi.org/10.1088/1367-2630/ac86e7}}.

\bibitem[Wootters(1998)]{Wootters1998}
Wootters, W.K.
\newblock {Entanglement of formation of an arbitrary state of two qubits}.
\newblock {\em Phys. Rev. Lett.} {\bf 1998}, {\em 80},~2245--2248,
  \href{http://xxx.lanl.gov/abs/9709029}{{\normalfont
  [arXiv:quant-ph/9709029]}}.
\newblock {\url{https://doi.org/10.1103/PhysRevLett.80.2245}}.

\bibitem[Jozsa(1994)]{Jozsa1994}
Jozsa, R.
\newblock {Fidelity for Mixed Quantum States}.
\newblock {\em J. Mod. Opt.} {\bf 1994}, {\em 41},~2315--2323.
\newblock {\url{https://doi.org/10.1080/09500349414552171}}.

\bibitem[Bellomo \em{et~al.}(2007)Bellomo, {Lo Franco}, and
  Compagno]{Bellomo2007}
Bellomo, B.; {Lo Franco}, R.; Compagno, G.
\newblock {Non-Markovian Effects on the Dynamics of Entanglement}.
\newblock {\em Phys. Rev. Lett.} {\bf 2007}, {\em 99},~1--4.
\newblock {\url{https://doi.org/10.1103/PhysRevLett.99.160502}}.

\bibitem[Bose(2007)]{Bose2007}
Bose, S.
\newblock {Quantum communication through spin chain dynamics: an introductory
  overview}.
\newblock {\em Contemp. Phys.} {\bf 2007}, {\em 48},~13--30.
\newblock {\url{https://doi.org/10.1080/00107510701342313}}.

\bibitem[Pan \em{et~al.}(1998)Pan, Bouwmeester, Weinfurter, and
  Zeilinger]{Pan1998}
Pan, J.W.; Bouwmeester, D.; Weinfurter, H.; Zeilinger, A.
\newblock {Experimental Entanglement Swapping: Entangling Photons That Never
  Interacted}.
\newblock {\em Phys. Rev. Lett.} {\bf 1998}, {\em 80},~3891--3894.
\newblock {\url{https://doi.org/10.1103/PhysRevLett.80.3891}}.

\bibitem[Apollaro \em{et~al.}(2015)Apollaro, Lorenzo, Sindona, Paganelli,
  Giorgi, and Plastina]{Apollaro2015}
Apollaro, T.J.G.; Lorenzo, S.; Sindona, A.; Paganelli, S.; Giorgi, G.L.;
  Plastina, F.
\newblock {Many-qubit quantum state transfer via spin chains}.
\newblock {\em Phys. Scr.} {\bf 2015}, {\em T165},~014036.
\newblock {\url{https://doi.org/10.1088/0031-8949/2015/T165/014036}}.

\bibitem[Ac{\'{i}}n \em{et~al.}(2000)Ac{\'{i}}n, Andrianov, Costa, Jan{\'{e}},
  Latorre, and Tarrach]{Acin2000}
Ac{\'{i}}n, A.; Andrianov, A.; Costa, L.; Jan{\'{e}}, E.; Latorre, J.I.;
  Tarrach, R.
\newblock {Generalized Schmidt Decomposition and Classification of
  Three-Quantum-Bit States}.
\newblock {\em Phys. Rev. Lett.} {\bf 2000}, {\em 85},~1560--1563,
  \href{http://xxx.lanl.gov/abs/0003050}{{\normalfont
  [arXiv:quant-ph/0003050]}}.
\newblock {\url{https://doi.org/10.1103/PhysRevLett.85.1560}}.

\bibitem[Enr{\'{i}}quez \em{et~al.}(2018)Enr{\'{i}}quez, Delgado, and
  {\.{Z}}yczkowski]{Enriquez2018a}
Enr{\'{i}}quez, M.; Delgado, F.; {\.{Z}}yczkowski, K.
\newblock {Entanglement of Three-Qubit Random Pure States}.
\newblock {\em Entropy} {\bf 2018}, {\em 20},~745.
\newblock {\url{https://doi.org/10.3390/e20100745}}.

\bibitem[D{\"{u}}r \em{et~al.}(2000)D{\"{u}}r, Vidal, and Cirac]{Du2000}
D{\"{u}}r, W.; Vidal, G.; Cirac, J.I.
\newblock {Three qubits can be entangled in two inequivalent ways}.
\newblock {\em Phys. Rev. A} {\bf 2000}, {\em 62},~062314.
\newblock {\url{https://doi.org/10.1103/PhysRevA.62.062314}}.

\bibitem[Coffman \em{et~al.}(1999)Coffman, Kundu, and Wootters]{Coffman2000}
Coffman, V.; Kundu, J.; Wootters, W.K.
\newblock {Distributed Entanglement}.
\newblock {\em Phys. Rev. A} {\bf 1999}, {\em 61},~052306,
  \href{http://xxx.lanl.gov/abs/9907047}{{\normalfont
  [arXiv:quant-ph/9907047]}}.
\newblock {\url{https://doi.org/10.1103/PhysRevA.61.052306}}.

\bibitem[Ma \em{et~al.}(2011)Ma, Chen, Chen, Spengler, Gabriel, and
  Huber]{Ma2011a}
Ma, Z.H.; Chen, Z.H.; Chen, J.L.; Spengler, C.; Gabriel, A.; Huber, M.
\newblock {Measure of genuine multipartite entanglement with computable lower
  bounds}.
\newblock {\em Phys. Rev. A} {\bf 2011}, {\em 83},~062325,
  \href{http://xxx.lanl.gov/abs/1101.2001}{{\normalfont [1101.2001]}}.
\newblock {\url{https://doi.org/10.1103/PhysRevA.83.062325}}.

\bibitem[Verstraete \em{et~al.}(2002)Verstraete, Dehaene, {De Moor}, and
  Verschelde]{Verstraete2002}
Verstraete, F.; Dehaene, J.; {De Moor}, B.; Verschelde, H.
\newblock {Four qubits can be entangled in nine different ways}.
\newblock {\em Phys. Rev. A} {\bf 2002}, {\em 65},~052112,
  \href{http://xxx.lanl.gov/abs/0109033}{{\normalfont
  [arXiv:quant-ph/0109033]}}.
\newblock {\url{https://doi.org/10.1103/PhysRevA.65.052112}}.

\bibitem[Osterloh and Siewert(2010)]{Osterloh2010}
Osterloh, A.; Siewert, J.
\newblock {The invariant-comb approach and its relation to the balancedness of
  multipartite entangled states}.
\newblock {\em New J. Phys.} {\bf 2010}, {\em 12}.
\newblock {\url{https://doi.org/10.1088/1367-2630/12/7/075025}}.

\end{thebibliography}
\end{document}